 \documentclass[prl,amsmath,amssymb,twocolumn,superscriptaddress]{revtex4}
\usepackage{cases}
\usepackage{amsmath}
\usepackage{amssymb}
\usepackage{amsfonts}
\usepackage{amssymb}
\usepackage{dcolumn}
\usepackage{bm}
\usepackage{bbm}
\usepackage{graphicx}
\usepackage{xcolor}
\usepackage{array}
\usepackage{subfigure}
\newcommand{\mJ}{{\mathcal J}}

\newcommand{\mF}{{\mathcal F}}

\newcommand{\tr}{{\rm Tr}}

\begin{document}
\title{Robertson-Schr\"odinger Uncertainty Relation  Refined by Skew Information }
\author{Sixia Yu}
\affiliation{Centre for quantum technologies, National University of Singapore, 3 Science Drive 2, Singapore 117543, Singapore}
\affiliation{
Hefei National Laboratory for Physical Sciences at Microscale and Department of Modern Physics,  University of Science and Technology of China, Hefei, Anhui 230026, China}
\author{C.H. Oh}
\affiliation{Centre for quantum technologies, National University of Singapore, 3 Science Drive 2, Singapore 117543, Singapore}
\affiliation{Physics department, National University of Singapore, 3 Science Drive 2, Singapore 117543, Singapore}
\begin{abstract}
We report a refinement of Robertson-Schr\"odinger uncertainty relation via Wigner-Yanase skew information. Besides the well known quantum uncertainty arising from the noncommutativity of observables, there is classical uncertainty arising from the mixedness of the states that is quantified by the difference between the variance and the skew information. Our refined uncertainty relation for canonical observables is saturated by all the Gaussian states, pure or mixed, and thus provides an alternative measure for  the non-Gaussianity of quantum states. Generalizations to the case of metric adjusted skew information are presented, unifying and refining most of previous results.
\end{abstract}
\maketitle
One of fundamental feature of quantum theory is Heisenberg's uncertainty principle, which states that {\it canonically conjugated observables can only be simultaneously determined with a characteristic uncertainty}  \cite{hur}.  The mere existence of incompatible observables already leads to  the quantum contextuality \cite{KS}, another profound nonclassical feature of quantum theory. The tradeoffs between the accuracies of {\it measuring} or {\it preparing} different observables are expressed via various kinds of uncertainty relations, providing also a quantification of Bohr's complementarity principle. Practically, uncertainty relations  find numerous applications ranging from setting the fundamental limit of the accuracy of estimating some unknown parameters, as in the quantum metrology \cite{qmetro}, to the detection of quantum entanglement \cite{ed,edyu}.

Soon after Heisenberg's original qualitative derivation of uncertainty relation Kennard and  Weyl \cite{kw} proposed  
the exact mathematical formulation of uncertainty relation for {\it preparation}. Interestingly an exact formulation of uncertainty relations for {\it measurement} was obtained only recently by Werner \cite{werner} via a joint-measurement approach and by Ozawa \cite{ozawa} via a measurement-disturbance approach.  Schr\"odinger \cite{sur} refined the Heisenberg uncertainty relation by the correlations of two observables and Robertson \cite{rur} further generalized to the case of more than two observables. 
For $n$ observables $\{X_k\}_{k=1}^n$ Robertson-Schr\"odinger (RS) uncertainty relation reads\begin{equation}\label{rsur}
 \left|\sigma_X\right|\ge\left|i\delta_X\right|,
\end{equation}
where  $\sigma_X$ is the covariance matrix and  $\delta_X$ is the  matrix formed by commutators with matrix elements 
\begin{subequations}
\begin{eqnarray}
[[\sigma_X]]_{kj}&=&\frac12\langle X_k X_j+X_jX_k\rangle_\varrho-\langle X_k\rangle_\varrho\langle X_j\rangle_\varrho,\\{}
[[\delta_X]]_{kj}&=&\frac i2\langle[X_k,X_j]\rangle_\varrho,
\end{eqnarray}
\end{subequations}
for $j,k=1,2,\ldots,n$ with $\langle O\rangle_\varrho=\tr\varrho O$ being the expectation value of an observable $O$ in the state $\varrho$. Here we have denoted by $|A|$ the determinant of a square matrix $A$.
Both RS and Heisenberg uncertainty relations can be saturated. In the case of  canonical observables, e.g., positions and momenta, Heisenberg's uncertainty relation is saturated by the coherent states and a restrict family of squeezed states while the RS uncertainty relation Eq.(\ref{rsur}) is saturated by all pure Gaussian states.  The stronger the uncertainty relation, the larger is the family of the minimal uncertainty states.

Many efforts have been made to further refine RS uncertainty relation, e.g., by using skew information.  Based on several nice properties such as convexity and additivity,  Wigner and Yanase (WY) \cite{wy63} introduced their skew information 
 \begin{equation}
I_\varrho(X^\dagger,X)=-\frac12\tr[\sqrt\varrho,X^\dagger][\sqrt\varrho,X],
 \end{equation} 
to quantify the information content of a quantum mechanical state $\varrho$ with respect to observables not commuting with (i.e., skew to) the conserved quantity
$X$. Being a measure for the noncommutativity between
a state $\varrho$ and an observable $X$, the skew information provides a measure of quantum uncertainty of $X$ in the state $\varrho$ and was used by Luo to derive a refinement of Heisenberg's uncertainty relation for mixed state \cite{luo03,luo05}. Furuichi \cite{F08} presented a refinement, taking into account the correlations, that is independent of RS uncertainty relation. Park \cite{park05} derived a refinement of Schr\"odinger's uncertainty for two observables, which can be saturated by a mixed state. However Park's approach is somewhat complicated and cannot be easily  generalized to more than three observables.

In this Letter we report a genuine refinement of RS uncertainty relation for $n$ observables by the skew information. In terms of the skew information matrix  $I_X$ with matrix elements $
[ [I_X]]_{kj}=I_\varrho(X_k,X_j)$ and $c_X=\sigma_X-I_X$, our refined RS uncertainty relation reads
\begin{eqnarray}\label{orur}
\left|\sigma_X+c_X|\cdot|\sigma_X-c_X\right|\ge|\delta_X|^2.
\end{eqnarray}
The nontrivial refinement over the RS uncertainty relation Eq.(\ref{rsur}) of the above uncertainty relation  is shown explicitly by its two weaker versions as below
\begin{equation}\label{orur2}
\left|\sigma_{X}\right|^{\frac2n}-\left|\delta_{X}\right|^{\frac2n}\ge\left((\left| \sigma_{X}\right|^{\frac 1n}-\left| I_X\right|^{\frac1n}\right)^{2}\ge\left|c_X\right|^{\frac 2n}.
\end{equation}
The additional uncertainty $|c_X|^{1/n}$ can be regarded as classical because it vanishes for pure states and is a concave function of the state. Then we shall show that our refined RS uncertainty relation Eq.(\ref{orur}) for canonical observables is saturated by all the Gaussian states, pure or mixed, providing an alternative measure for the non-Gaussianity of a quantum state. Finally  our refined uncertainty relation is generalized to the case of metric adjusted skew information, unifying and refining most of previous results.

Our main result is based on a simple observation that leads to RS uncertainty relation \cite{rur} as well as a simple derivation \cite{ghp09} of the dynamic uncertainty relation \cite{dyur}, with a special case being $\sigma_X\ge I_X$ \cite{luo2}. From a set of $n$ observables $\{X_k\}$, by denoting  $X_k^\prime=X_k-\langle X_k\rangle_\varrho$, we introduce a set of $2n$ operators 
\begin{equation}
Y_{k\pm}=\frac {\sqrt\varrho X_k^\prime\pm X_k^\prime\sqrt\varrho}{\sqrt2}:=\frac1{\sqrt2}[\sqrt\varrho,X_k^\prime]_\pm
\end{equation}
with $\varrho$ being a given state.
This set of operators has been used by Park to derive a refined RS uncertainty relation for two observables \cite{park05}.
Let $L_X$ denote the $2n\times 2n$ matrix whose matrix elements are $[[L_X]]_{k\mu,j\nu}=\tr Y_{k\mu}^\dagger Y_{j\nu}$ for $j,k=1,2,\ldots,n$ and $\mu,\nu=\pm$. 
Explicitly we have
\begin{subequations}
\begin{eqnarray}
\tr(Y_{k\pm}^\dagger Y_{j\pm})&=&\pm\frac12\tr[\sqrt\varrho,X_k^\prime]_\pm[\sqrt\varrho,X_j^\prime]_\pm,\\{}
\tr(Y_{k+}^\dagger Y_{j-})&=&-\frac12\langle[X_k,X_j]\rangle_\varrho=i[[\delta_X]]_{kj}.
\end{eqnarray}
\end{subequations}
The simple observation reads $L_X\ge0$ since $L_X$ can be regarded as the Gram matrix of $2n$ operators $Y_{k\mu}$ with respect to the inner product $\tr X^\dagger Y$.  When arranged in a block form, with each block matrix of size $n\times n$, the condition $L_X\ge0$ becomes
\begin{equation}\label{Lx}
L_X=\left(\begin{array}{cc} \sigma_X+c_X & i\delta_X\\
-i\delta_X&\sigma_X-c_X\end{array}\right)\ge0
\end{equation}
 which is the matrix form of our refined RS uncertainty relation. Using  Schur complement condition for positive semidefinite, we obtain 
\begin{equation}\label{lx2}
\sigma_X+c_X\ge\delta_X\frac1{\sigma_X-c_X}\delta_X.
\end{equation} 
If $\sigma_X-c_X$ has some zero eigenvalues we have only to understand its inverse appearing in Eq.(\ref{lx2}) as being defined in its range, which contains the range of $\delta_X$ as $L_X\ge0$.

Starting from the matrix form one can obtain various scalar uncertainty relations expressed via various characteristics of the positive semidefinite matrix $L_X$, as proposed by Trifonov and Donev \cite{tr}. For example  all the principal minors of $L_X$ must be nonnegative. As a special case, by taking the determinants of both sides of Eq.(\ref{lx2}) we obtain immediately our refined RS uncertainty relation Eq.(\ref{orur}). 
  In the case of two observables stronger scaler uncertainty relations are possible. In fact the matrix form  Eq.(\ref{lx2})  can be equivalently characterized by the following set of scalar uncertainty relations (see Appendix 1)
  \begin{subequations}
\begin{eqnarray}\label{21}
|\sigma_X|-|c_X|-\sqrt{(|\sigma_X|-|c_X|)^2-|L_X^+L_X^-|}\ge\delta^2,\\
\frac{L_a^+}{L_a^-}|L_X^-|\ge\delta^2 \quad (a=1,2),\label{22}
\end{eqnarray} 
\end{subequations}
where we have denoted $\delta=\langle[X_1,X_2]\rangle_\varrho/2$ and $L_X^{\pm}=\sigma_X\pm c_X$ are $2\times 2$ matrix with matrix elements denoted by $L_a^\pm=[[L_X^\pm]]_{aa}$ with $a=1,2$ and $L_{12}^\pm=[[L_X^\pm]]_{12}$. We note that uncertainty relation Eq.(\ref{21}) is stronger than the uncertainty relation Eq.(\ref{orur}) for two observables.

To derive the weaker versions Eq.(\ref{orur2}) of our refined RS uncertainty relation  we need to employ the Minkowski's  inequality  for the determinants of positive semidefinite matrices:
 $ |A|^{\frac 1n}-|B|^{\frac 1n}\ge |A-B|^{\frac 1n}$ for two $n\times n$ Hermitian matrices $A\ge B\ge0$. The first inequality in Eq.(\ref{orur2}) is obtained by applying Minkowski's inequality for $A=2\sigma_X$  and $B=I_X$ together with our refined RS uncertainty relation Eq.(\ref{orur}). The second inequality  in Eq.(\ref{orur2}) is obtained by applying Minkowski's inequality one more time for $A=\sigma_X$ and $B=I_X$. 

From the weaker but suggestive versions Eq.(\ref{orur2}), especially the second one, it is tempting to introduce a two-dimensional uncertainty vector $(|\delta_X|^{1/n},|c_X|^{1/n})$ whose length provides the lower bound of the variance. The first component $|\delta_X|^{1/n}$ is quantum uncertainty since it arises from the non commutativity among observables. The second component $|c_X|^{1/n}$ can be regarded as a kind of classical uncertainty for two reasons. First, it  comes from the mixing of the quantum states and vanishes for pure state. Second, it is a concave function of the state $\varrho$. This is because the WY skew information $I_\varrho(X^\dagger,X)$ is a convex function of the state $\varrho$ so that the skew information matrix $I_X$ is a convex matrix function of $\varrho$. As a result the classical uncertainty matrix $c_X$ is a concave matrix function of $\varrho$ and, due to Minkowski's inequality, the classical uncertainty $|c_X|^{1/n}$ is a concave function of $\varrho$. This means that the more mixing of the state the larger is the classical uncertainty.

One of the main reasons why the quantity $|c_X|^{1/n}$ can be regarded as classical uncertainty is that even if those $n$ observables are commuting there is still a nontrivial lower bound for the variance that is arising from the mixedness of the quantum states, i.e., the uncertainty of which pure states. Also in the case of an odd number of observables the quantum uncertainty, as given by the determinants of the commutator matrix, also vanishes and the classical uncertainty provides a nontrivial bound for mixed states, similar to the case of dynamical uncertainty relation as noticed in \cite{ghp09}.

Luo \cite{luo05,luo3} also advocated a separation of the classical and quantum uncertainties and obtained a refinement of Heisenberg's uncertainty relation for two observables \cite{luo05}. Taking $U_{X_a}^2=\sigma_{X_a}^2-c_{X_a}^2$ as a measure of the quantum uncertainty  for each observable $X_{a}$ with $a=1,2$,   Luo managed to prove that $U_{X_1}U_{X_2}\ge \delta^2$, which improves Heisenberg uncertainty relation $\sigma_{X_1}\sigma_{X_2}\ge\delta^2$ since $\sigma_X\ge U_X$. 
Our refined RS uncertainty relation Eq.(\ref{orur}) improves that of Luo considering Cauchy's inequality
\begin{equation}
U_{X_1}U_{X_2}\ge\sqrt{\left|L_X^+L_X^-\right|}+|L_{12}^+L_{12}^-|,
\end{equation}
and $|L_X^\pm|=L_{1}^\pm L_{2}^\pm-(L_{12}^\pm)^2$. Furuichi improved Luo's result by showing $U_{X_1}U_{X_2}\ge \delta^2+(L_{12}^-)^2$ \cite{F08}, which can be further refined by our uncertainty relations Eq.(\ref{22}) (see Appendix 1). We believe that (without a proof)  Park's refinment \cite{park05} can also be derived from uncertainty relations Eq.(\ref{21}) and Eq.(\ref{22}).

Our refined RS uncertainty relation Eq.(\ref{orur}) can also be saturated. Denoting $\Delta_G:=|L_X^+L_X^-|-|\delta_X|^2$ with $L_X^\pm=\sigma_X\pm c_X$ and from the inequality $\Delta_G\ge |L_X|$ it is clear that a necessary condition for our refined RS uncertainty relation Eq.(\ref{orur}) to be attained is that $L_X$ has some zero eigenvalues. That is to say $2n$ operators $Y_{k\mu}$ are linearly dependent, i.e., there exist $2n+1$ complex numbers $a_k,b_k,c$ such that $\sum_k (a_k \sqrt\varrho X_k+ b_k X_k\sqrt\varrho) = c\sqrt\varrho,$ which amounts to requiring that the state $\varrho$ generates a linear transformation among observables $X_k$, e.g., $\varrho X_i\varrho^{-1}$ is a linear combination of $X_i$. 
 
Consider an $n$-mode bosonic system or $n$ interacting quantum harmonic oscillators with their annihilation and creation operators, denoted collectively by 
\begin{equation}
\Lambda=(a^\dagger,a)=(a^\dagger_1,a^\dagger_2,\ldots,a^\dagger_n,a_1,a_2,\ldots,a_n),
\end{equation}  
satisfying $[a_j,a_k^\dagger]=\delta_{jk}$.
Let $\varrho\propto e^{-\beta H}$ be the thermal state of a most general quadratic Hamiltonian 
\begin{equation}\label{H}
 H=\frac 12\Lambda NJ \Lambda^T,\quad J=\left(\begin{array}{cc}0&I_n\\-I_n&0\end{array}\right)
\end{equation}
in which the transposition acts only on $2n\times 2n$ matrix without affecting the bosonic operators and $NJ$ is a $2n\times 2n$ symmetric matrix such that $H$ is Hermitian.

The correlation matrix $C_\Lambda$ of $2n$ operators $\Lambda$ in the thermal state $\varrho$, whose matrix elements are given by $[[C_\Lambda]]_{kj}=\tr\varrho\Lambda_k\Lambda_j$, can be readily calculated with the help of linear quantum transformation theory \cite{lqt}.
From the commutators $[\Lambda^T,\Lambda]=J^T$ it follows immediately $[H,\Lambda]=\Lambda N$ and  the identity $e^{-\beta H}\Lambda e^{\beta H}=\Lambda M$ with $M=e^{-\beta N}$, which is obtained by Heisenberg's equation of motion. Suppose that $M-I_{2n}$ is invertible and it follows $\tr\varrho\Lambda=0$. From identities 
\begin{subequations}
\begin{eqnarray}\label{ct}
\tr\varrho \Lambda_j\Lambda_k=\tr[[\Lambda M]]_j \varrho \Lambda_k=[[C_\Lambda M]]_{kj},\\
\label{cc}
\tr\sqrt\varrho\Lambda_j\sqrt\varrho\Lambda_k=\tr[[\Lambda \sqrt M]]_j \varrho \Lambda_k=[[C_\Lambda \sqrt M]]_{kj}.
\end{eqnarray}
\end{subequations}
it follows $C^T_\Lambda=C_\Lambda M$ and $c_{\Lambda}=C_\Lambda \sqrt M$. Considering $J=C^T_\Lambda-C_\Lambda=2\delta_\Lambda$ and $\sigma_\Lambda=(C^T_\Lambda+C_\Lambda)/2$ we obtain 
\begin{equation}
C_\Lambda=J\frac{I_{2n}}{M-I_{2n}},\quad\sigma_\Lambda\pm c_\Lambda=\frac12J\frac {\sqrt M\pm I_{2n}}{\sqrt M\mp I_{2n}},
\end{equation}
from which it follows $|(\sigma_\Lambda+ c_\Lambda)(\sigma_\Lambda- c_\Lambda)|=|\delta_\Lambda|^2$, which also holds in the case of singular $M-I_{2n}$ since we can always perturb slightly the parameters in $N$ to ensure that the new $M-I_{2n}$ is invertible.
Now we consider $2n$ canonical observables $X=(x,p)=\Lambda u$, e.g., positions and momenta in the case of harmonic oscillators, where 
\begin{equation}
u=\frac1{\sqrt2}\left(\begin{array}{cc}I_n&iI_n\\I_n&-i I_n\end{array}\right).
\end{equation}
It turns out that $\sigma_X=u^T\sigma_\Lambda u$, $c_X=u^Tc_\Lambda u$, and $\delta_X=u^T\delta_\Lambda u$ so that our refined uncertainty relation Eq.(\ref{orur}) for $2n$ canonical observables $X=\Lambda u$ is saturated by  the thermal states of all the quadratic Hamiltonians. 

On the other hand a Gaussian state is determined completely by its correlation matrix $C_\Lambda$, supposing $\tr\varrho\Lambda=0$. Provided that $C_\Lambda$ is invertible, the matrix $M^\prime=C^{-1}C^T$ belongs to symplectic group, i.e., satisfies $M^{\prime T}J M^\prime=J$. Thus there exists an element $N^\prime$ in the symplectic algebra  such that $M^\prime=e^{-N^\prime}$. The thermal state of the corresponding  quadratic Hamiltonian Eq.(\ref{H}) for $N^\prime$  at temperature $\beta=1$ has exactly the same correlation matrix $C_\Lambda$. By an argument of continuity, any Gaussian state can be approximated by the thermal state of a quadratic Hamiltonian so that all the Gaussian states, pure or mixed, saturate our refined RS uncertainty relation Eq.(\ref{orur}). Thus the nonzero difference $\Delta_G$ signals the non-Gaussianity of a quantum state. 

Let us now explore some generalizations of our refined RS uncertainty relation. For any bivariable function $g(x,y)$ and a given state $\varrho$ with eigensystem $\{\lambda_k, |\psi_k\rangle\}$ we introduce a generalized covariance matrix  $\sigma_X(g)$, called  $g$-covariance for short,  with matrix elements
\begin{subequations}
\begin{eqnarray}
[[\sigma_X({g})]]_{kj}=\tr X_k \mJ_\varrho^{g}(X_j),\\ \mJ_\varrho^g(Z)=\sum_{j,k} g(\lambda_j,\lambda_k)P_jZ P_k,\quad P_k=|\psi_k\rangle\langle\psi_k|.
\end{eqnarray}
\end{subequations}
 for a set of $n$ observables $\{X_k\}_{k=1}^n$. 
For examples the covariance matrix $\sigma_X$ corresponds to $g$-covariance $\sigma_X(g)$ with $g(x,y)=(x+y)/2$ while the commutator matrix $\delta_X$ corresponds to $g$-covariance $\sigma_X(\epsilon)$ with $\epsilon(x,y)=i(y-x)/2$  since  $\tr X_k \mJ_\varrho^{\epsilon}(X_j)=[[\delta_X]]_{kj}$. 
 It is clear that $\sigma_X(cg)=c\sigma_X(g)$  for any complex number $c$ and $\sigma_X(g_1+g_2)=\sigma_X(g_1)+\sigma(g_2)$. 
 
{\it Observation 1} If $g(x,y)\ge 0$ for $x,y\ge0$, since $\tr Z^\dagger\mJ_\varrho^g(Z)\ge 0$ for an arbitrary operator $Z$, then we have $\sigma_X({g})\ge0$.  As an immediate consequence we have $\sigma_X({g_1})-\sigma_X({g_2})=\sigma_X({g_1-g_2})\ge 0$ if  two bivariable functions satisfying $g_1(x,y)\ge g_2(x,y)$ for $x,y\ge0$.

{\it Observation 2} For two arbitrary bivariable functions $g_a(x,y)$ with $a=1,2$ and two operators $Y,Z$ we have $\tr\mJ^{g_1}_\varrho(Y)^\dagger\mJ^{g_2}_\varrho(Z)=\tr Y^\dagger\mJ^{g_1^*g_2}_\varrho(Z)$.  As the Gram matrix of $2n$ observables $Y_{ka}=\mJ_\varrho^{g_a}(X_k)$  with $k=1,2,\ldots, n$ and $a=1,2$ with respect to inner product $\tr Y^\dagger Z$, the $2n\times 2n$ matrix $L_X^g$ matrix  defined by $[[L_X^g]]_{ka,jb}=\tr Y_{ka}^\dagger Y_{jb}$ should be nonnegative, i.e.,
\begin{equation}\label{grur}
L_X^g=\left(\begin{array}{cc} \sigma_X(|g_1|^2) & \sigma_X({g_1^*g_2})\\
\sigma_X({g_2^*g_1})&\sigma_X(|g_2|^2)\end{array}\right)\ge0.
\end{equation}
This is a generalization of our refined uncertainty relation in matrix form Eq.(\ref{Lx}). As an immediate application, for any three functions $g_\pm(x,y)\ge0,g_0(x,y)$ satisfying $g_+g_-\ge |g_0|^2$ it holds 
\begin{equation}\label{urz}
|\sigma_X({g_+})|\cdot|\sigma_X({g_-})|\ge|\sigma_X({g_0})|^2
\end{equation} 
since $\sigma_X({g_+})\ge\sigma_X(|g_0|^2/g_-)$.  Considering two functions $a_x,b_x$ such that $g_\pm(x,y)=(a_x\pm a_y)(b_x\pm b_y)\ge0$ and $g_+g_-\ge g_0^2$ where $g_0=\mu(a_xb_y-a_yb_x)$ for some constant $\mu$, e.g., as given in \cite{ko11}, we have a modified commutator matrix $[[\sigma_X({g_0})]]_{kj}=\mu\tr a_\varrho b_\varrho[X_k,X_j]$.  The uncertainty relation Eq.(\ref{urz}) improves those Heisenberg type of uncertainty relations, e.g., as in \cite{ko11}, with a modified commutator matrix.

Immediate after its discovery, WY skew information was generalized by Dyson to a one-parametered family, called Wigner-Yanase-Dyson (WYD) skew information \cite{wy63}. Recently the skew information is further generalized by Hansen \cite{H06} to a most general family of skew information, called metric adjusted skew information or $f$-skew information, 
\begin{equation}
I_\varrho^f(X^\dagger,X)=\frac{f(0)}2\sum_{j,k}\frac{(\lambda_k-\lambda_j)^2}{\lambda_jf(\lambda_k/\lambda_j)}\tr P_kX^\dagger P_j X
\end{equation}
that is parametrized by the whole set $\mF$ of regular symmetric operator monotone functions $f(x)$.  A nonnegative function $f(x)$ for $x\ge0$  is  operator monotone if  $f(A)\le f(B)$ for any two Hermitian matrices satisfying $0\le A\le B$, symmetric if  $xf(1/x)=f(x)$,  normalized if  $f(1)=1$, and regular if $f(0)>0$ \cite{petz96}. 
 WYD skew information is  the $f$-skew information corresponding to 
\begin{equation}
f_{\alpha}(x)=\frac{\alpha(1-\alpha)(1-x)^2}{(1-x^\alpha)(1-x^{1-\alpha})},\quad 0<\alpha\le \frac 12.
\end{equation}
with $\alpha=1/2$ being the WY skew information. The $f$-skew information associated with $f_M(x)=(1+x)/2$ becomes the quantum Fisher information \cite{qfi}, up to some constant factor, in certain cases. The $f$-skew information matrix $I_X^f$, whose matrix elements are given by $[[I_X^f]]_{kj}=I_\varrho^f(X_k,X_j)$, can be regarded as a  $g$-covariance. In fact if we denote $m_f(x,y)=yf(x/y)$ for an arbitrary $f(x)$ then we have $I_X^f=\sigma_X({m_{f_*}})$ with $f_*(x)=f(0){(1-x)^2}/[2f(x)].$
 
As another application, for a regular $f\in \mF$ we take $g_1=\sqrt{m_f}$ and $g_2=\epsilon/g_1$ in our observation 2. In this case we have $\sigma_X(g_1^2)=\sigma_X({m_f})$, $\sigma_X({|g_2|^2})=I_X^f/[2f(0)]$, and $ \sigma_X(\epsilon)=\delta_X$ and from Eq.(\ref{grur}) it follows
\begin{equation}\label{g1}
|\sigma_X({m_f})|\cdot |I_X^f|\ge[2f(0)]^n|\delta_X|^2.
\end{equation} 
By denoting $\lambda_f=\min_{x\ge0}\left(1+x-f_*(x)\right)/[2f(x)]$, we have $2f_M(x)-f_*(x)\ge2 \lambda_f f(x)$ 
so that $2m_{f_M}-m_{f_*}\ge \lambda_fm_f$ and thus $2\sigma_X-I_X^f\ge2\lambda_f\sigma_X({m_f})$ due to observation 1. Thus we obtain
\begin{equation}\label{fur}
|\sigma_X-c_X^f|\cdot|\sigma_X+c_X^f|\ge[4\lambda_ff(0)]^n\left|\delta_X\right|^{2}
\end{equation}
with $c_X^f=\sigma_X-I_X^f$ being the metric adjusted classical uncertainty matrix.
For WYD skew information we have $\lambda_{f_\alpha}=1$ and in general $1-f(0)\le\lambda_f\le \min\{1,1/[4f(0)]\}$ (see Appendix 2). Generalized uncertainty relation Eq.(\ref{fur}) refines those results in \cite{Yanagi11,fy12}. By  taking $f(x)=f_{\alpha=\frac12}(x)$ the metric adjusted skew information $I_X^f$ becomes the WY skew information and both two generalizations above reduce to our refined RS uncertainty relation Eq.(\ref{orur}). 

In summary we have derived such a strong refinement of RS uncertainty relation Eq.(\ref{orur}) that all the Gaussian states, pure or mixed, become minimal uncertainty states.  The nonzero difference $\Delta_G$  between two sides of our refined RS uncertainty relation Eq.(\ref{orur}) provides therefore a natural measure for non-Gaussianity of quantum states. A classical uncertainty that arises from the mixing of pure states is identified and quantified by the difference between the variance and WY skew information. Generalizations Eq.(\ref{g1}) and Eq.(\ref{fur}) to the metric adjusted skew informations are also presented and corresponding minimal uncertainty states  may have potential applications in quantum optics and quantum computational tasks, like Gaussian states. Also the applications in entanglement detection as well as quantum metrology may be expected. At last our refined uncertainty relation should be helpful to sharpen, e.g., Ozawa's uncertainty relation for measurement and disturbances \cite{ozawa}, which has been tested experimentally \cite{exps}.

This work is supported by National Research Foundation
and Ministry of Education, Singapore (Grant No.
WBS: R-710-000-008-271) and NSF of China (Grant No.
11075227).

\setcounter{equation}{0}
\renewcommand{\theequation}{A.\arabic{equation}}

{\it Appendix 1 Proof of Eq.(\ref{21}) and Eq.(\ref{22}). --- } In the case of two observables $\delta_X$ and $L_X^\pm=\sigma_X\pm c_X$ are two by two matrices. Since $\delta_X=\delta\sigma_y$ with $\delta=\langle [X_1,X_2]\rangle_\varrho/2$ and $\sigma_y$ being the second Pauli matrix and for an arbitrary two by two matrix $U$ it holds $\sigma_y U^T\sigma_y U=|U|$, the matrix form of uncertainty relation Eq.(\ref{lx2}) becomes $|L_X^-|L_X^+- \delta^2 L_X^-\ge0$, whose diagonal elements are exactly Eq.(\ref{22}) and whose determinant leads to inequality 
\begin{equation}
\delta^4-2A\delta^2+B\ge0
\end{equation}
with $A=|\sigma_X|-|c_X|$ and $B=|L_X^+L_X^-|$. As a quadratic function of $\delta^2$, the above inequality leads to either Eq.(\ref{21}) or $A+\sqrt{A^2-B}\le \delta^2$. However the second alternative  is impossible because the refined RS uncertainty relation Eq.(\ref{orur}) for $n=2$ leads to $A\ge\delta^2$. Note that Eq.(\ref{22}) can be rewritten as 
\begin{eqnarray}
L_1^+L_2^-&\ge& \delta^2+\frac{L_1^+}{L_1^-}(L_{12}^-)^2\\
L_2^+L_1^-&\ge& \delta^2+\frac{L_2^+}{L_2^-}(L_{12}^-)^2
\end{eqnarray}
from which it follows  
\begin{equation}
U_{X_1}U_{X_2}\ge\delta^2+\sqrt{\frac{L_1^+L_2^+}{L_1^-L_2^-}}(L_{12}^-)^2
\end{equation}
which refines Furuichi's result $U_{X_1}U_{X_2}\ge\delta^2+(L_{12}^-)^2$ \cite{F08} because $L_a^+\ge L_a^-$ due to the fact that $c_X\ge 0$ and $L_a^\pm=[[L_X^\pm]]_{aa}=\sigma_{X_a}\pm c_{X_a}$.

{\it Appendix 2 Range of $\lambda_f$. --- }
Here we shall determine the range of $\lambda_f=\min_{x\ge 0}F(x)$ with
\begin{equation}
F(x)=\frac1{2f(x)}\left(1+x-\frac{f(0)(1-x)^2}{2f(x)}\right)
\end{equation}
for an arbitrary normalized symmetric operator  monotone function $f(x)$. We have the upper bound $\lambda_f\le \min\{1,1/[4f(0)]\}$ because $F(1)=1$ and $F(0)=1/[4f(0)]$. To get the lower bound, we let $x_0\ge0$ achieve the minimum of $F(x)$, i.e., $\lambda_f=F(x_0)$. If we denote $z=(1+x_0)f(0)/[2f(x_0)]$, since $(1-x_0)^2\le (1+x_0)^2$, then we have $\lambda_f\ge z(1-z)/f(0)$. On the one hand, since $f\in \mF$, we have $(1+x_0)/2\ge f(x_0)$, i.e, we have $ 1/2\ge z$, so that the quadratic function $z(1-z)$ is increasing. On the other hand  
since $f(x)$ is concave we have $f^\prime(x)\ge f^\prime(\infty)=\lim_{x\to \infty}f(x)/x=\lim_{x\to \infty}f(1/x)=f(0)$. As a consequence $f(x)-f(1)\ge f(0)(x-1)$ for $x\ge 1$. For $x\le 1$ we have $1/x\ge 1$ and $f(1/x)\ge 1+f(0)(1-x)/x$ which, together with $f(x)=xf(1/x)$, leads to $f(x)\ge x+f(0)(1-x)$. Thus we obtain  
\begin{equation}
f(x)\ge \frac{1+x}2-\frac{1-2f(0)}2|1-x|\ge f(0)(1+x).
\end{equation}
As a result  we obtain $z\ge f(0)$ so that 
$\lambda_f\ge 1-f(0)=\lambda_f^*$. We note that 
 $\lambda_f^*\ge f(0)$ as $f(0)\le 1/2$. In general we have $F^\prime(1)=0$, i.e., $F(1)$ is always a local extremal point. Numerical  evidences show that  $\lambda_f=1$ whenever $4f(0)\le 1$ and we conjecture that $\lambda_f=\min\{1,1/[4f(0)]\}$ always holds.
For the monotone operator function $f_{\alpha}(x)$ corresponding to Wigner-Yanase-Dyson skew information it holds $\lambda_{f_\alpha}=1$ because $\lambda_{f_\alpha}\ge 1$ follows from the inequality 
\begin{equation}
|x^{\alpha}-x^{1-\alpha}|\le(1- 2\alpha)|1-x|,\quad 0<\alpha\le \frac12.
\end{equation}

For those $f$-skew informations corresponding to monotone functions $f(x)$ with  $\lambda_f=1/[4f(0)]$, or equivalently $f(x)\le f(0)(1+\sqrt x)^2$, e.g., $f_M(x)=(1+x)/2$ and $f_{1/2}(x)=(1+\sqrt x)^2/4$, the uncertainty relation Eq.(\ref{fur}) are refinements of RS uncertainty relation. However the one given in Eq.(\ref{orur}) corresponding to $f_{\alpha=1/2}(x)$ is the strongest. This is because $I_X\le I_X^{f}$ and $c_X\ge c_X^{f}$ which means that $r_k\ge r_k^f$ with $r_k^{(f)}$ being the eigenvalues of $\frac1{\sqrt{\sigma_X}}c^{(f)}\frac1{\sqrt{\sigma_X}}$ arranged in decreasing order and thus
\begin{eqnarray}
&&|\sigma_X-c_X^f|\cdot|\sigma_X-c_X^f|
=|\sigma_X|^2\prod_{k}(1-(r_k^f)^2)\cr
&\ge&|\sigma_X|^2\prod_{k}(1-r_k^2)
=|\sigma_X-c_X|\cdot|\sigma_X-c_X|.
\end{eqnarray}
For those $f$-skew informations corresponding to monotone functions $f(x)$ with  $\lambda_f\not =1/[4f(0)]$ the uncertainty relation Eq.(\ref{fur}) may be independent of RS uncertainty relation.

\newpage
\end{document}